\documentclass{aa} 

\begin{document}

\title{Physical parameters and emission mechanism in Gamma-Ray 
Bursts} 

\author{E.V. Derishev$^{1}$, V.V. Kocharovsky$^{1,2}$ \and Vl.V. 
Kocharovsky$^{1}$}

\offprints{E.V. Derishev} 

\institute{$^{1}$Institute of Applied Physics, Russian Academy of Science 
\\
46 Ulyanov st., 603950 Nizhny Novgorod, Russia\\
$^2$Dept. of Physics, Texas A\&M University, College Station,
TX 77843-4242, USA\\
e-mail: derishev@appl.sci-nnov.ru, kochar@appl.sci-nnov.ru}


\titlerunning{Physical parameters and emission mechanism in GRBs}

\authorrunning{Derishev, Kocharovsky \& Kocharovsky}

\abstract{
Detailed information on the physical
parameters in the sources of cosmological Gamma-Ray Bursts (GRBs)
is obtained from few plausible assumptions consistent with
observations. We consider monoenergetic injection of electrons 
and let them cool self-consistently, taking into account 
Klein-Nishina cut-off in electron-photon scattering.
The general requirements posed by the assumptions on the
emission mechanism in GRBs are formulated. It is found that the
observed radiation in the sub-MeV energy range is generated by the
synchrotron emission mechanism, though about
ten per cent of the total GRB energy should be converted via the
inverse Compton (IC) process into the ultra-hard spectral domain 
(above 100 GeV). We estimate the magnetic field strength in the 
emitting region, the Lorentz factor of accelerated electrons, and the 
typical energy of IC photons. 
We show that there is a synchrotron-self-Compton constraint 
which limits the parameter space available for GRBs that are 
radiatively efficient in the sub-MeV domain. This concept is 
analogous to the line-of-death relation existing for pulsars and 
allows us to derive the lower limits on both GRB duration and the 
timescale of GRB variability.  The upper limit on the Lorentz factor 
of GRB fireballs is also found.  We demonstrate that steady-state 
electron distribution consistent with the Compton losses may produce 
different spectral indices, e.g., 3/4 as opposed to the figure 1/2 
widely discussed in the literature.  It is suggested that the changes 
in the decline rate observed in the lightcurves of several GRB 
afterglows may be due to either a transition to efficient IC cooling 
or the time evolution of Klein-Nishina and/or Compton spectral 
breaks, which are the general features of self-consistent electron 
distribution.  
}
\maketitle

\section{Introduction}

At present, there is a tendency to divide the problem of Gamma-Ray
Bursts into two separate problems: the nature of the central
engine and the origin of GRB emission. While the former question is
still far from being settled, the multi-wavelength observations
(e.g., Frail et al. 1997; Metzger et al. 1997; Reichart 1997; Waxman
1997; see also Piran 1999 for a review), which become available in 
the past few years, give strong support to the fireball shock model 
(Rees \& M\'{e}sz\'{a}ros 1992, 1994). In its most general form, this 
model considers GRB emission as the result of energy dissipation in 
relativistic shock waves and assumes that all the energy is initially 
accumulated in the form of ultrarelativistic outflow (fireball) with 
a Lorentz factor $\Gamma$ greater than 100 (Baring \& Harding 1995).  
The fraction of energy carried directly by the emission from the 
fireball photosphere is small unless $\Gamma \ga 10^5$ (Derishev, 
Kocharovsky \& Kocharovsky 1999).

The fireball model became widespread and almost undisputed, but in 
its details there is no agreement among theorists. Even the principal 
question about the main emission mechanism has not yet received 
a definite answer, although synchrotron and inverse Compton emission 
are the favourites (e.g., Paczy\'{n}ski \& Rhoads 1993; 
M\'{e}sz\'{a}ros, Laguna \& Rees 1993; Sari, Piran \& Narayan 1998). 
The existing uncertainty complicates the correct interpretation of 
the growing number of dissimilar multi-wavelength observations.

In this paper we derive constraints on the physical parameters in the 
GRB emitting region and present new arguments in favour of the 
synchrotron emission mechanism. There are two arguments widely 
discussed in the literature in this connection. First, the observed 
polarization in afterglow emission (Covino {\it et al.} 1999; Wijers 
{\it et al.} 1999) is thought to be indicative of synchrotron 
mechanism. Second, it was argued that the electron cooling time is 
too long to be compatible with high radiative efficiency if one 
assumes self-Compton mechanism as the main source of sub-Mev 
radiation (Piran 1999). Both arguments are not very restrictive 
because: a) polarization may be produced by inverse Compton 
scattering as well if the radiation background and/or electron 
distribution is anisotropic, b) the strongest limit on the bulk 
Lorentz factor imposed by cooling time considerations is for the 
external shock model, $\Gamma < 300-400$, which is still above the 
generally accepted lower limit of $\sim 100$.  In this paper 
we show that the additional requirement of high efficiency in 
sub-MeV gamma-rays relative to other spectral domains pushes the 
upper limit on $\Gamma$ to an unacceptable value of $\sim 25$ unless 
the sub-MeV radiation is generated by synchrotron rather than 
by a self-Compton mechanism.

There is also an argument 
against a synchrotron-self-Compton model: it is sometimes 
considered as insufficiently flexible to account for different 
low-energy spectral indices provided the electrons form a cooling 
distribution. Below we show that this argument is rather  weak 
since a broad range of spectral indices may be accommodated within 
the framework of synchrotron-self-Compton model as long as the 
cooling distribution is calculated consistently with {\it total} 
losses. 

We introduce the minimum number of 
{\it a priori} assumptions and avoid the considerations which would 
require the electron acceleration mechanism to be specified. All the 
assumptions used are explicitly listed in the second section, 
followed by a brief discussion and a comparison with other approaches 
frequently used in the literature. The basic analysis is presented 
in the third section, and the implications are discussed in the 
forth section.

For numerical estimates we take GRB energy $E_{\rm GRB} =
10^{52}$~erg and duration $t_{\rm GRB} = 10$~s. All other values are
measured in the shock comoving frame unless the opposite is
stated. It should be noted that for two bursts, GRB971214 and 
GRB990123, the estimated value of energy release approaches 
$10^{54}$~ergs assuming isotropic emission. Since it is very 
divergent from the average value ($10^{51}-10^{52}$~ergs) inferred 
from the analysis of the logN-logS curve (Reichart \& 
M\'{e}sz\'{a}ros 1997), these bursts may be of different origin.

\section{Model assumptions}

{\it 1. Most GRB emission is in the form of soft
gamma-rays, i.e., below several MeV.} -- It would naturally explain
why GRB phenomenon was first detected in this spectral range.
The observation of optical emission from GRB990123 (Akerlof et al.
1999), made simultaneously with the burst of gamma-rays,
confirms our first assumption. However, observations do not exclude
the possibility that the main part of GRB emission may be
concentrated in photons with energies greater than 1 TeV, and hence
escapes detection because photons in this energy range are strongly
absorbed by infrared background radiation (Primack et al.
1999).

{\it 2. A major part of the fireball kinetic energy is converted into
electromagnetic radiation during the main GRB pulse.} -- GRB
afterglows decline faster than $t^{-1}$ and do not contribute
significantly to the total energy budget
(Galama et al. 1998).  However, it is possible that the adiabatic 
regime of the shock deceleration persists for a very long time, so 
that the bulk of GRB energy has never been observed. Another reason 
to make the second assumption is not to increase the energy 
requirements, which are already rather restrictive unless the 
fireball forms a narrow jet (e.g., M\'{e}sz\'{a}ros, Rees \& Wijers 
1999).

{\it 3. GRB emission is generated by relativistic electrons and
positrons, which are continuously injected with the initial Lorentz 
factor $\gamma_{\rm p}$. Acceleration of an electron takes a very 
short time compared to the shock expansion timescale and electron 
cooling timescale.} -- This same assumption implies that we refer to 
the synchrotron-self-Compton model of GRB emission. Comparison with 
the other emission models favours the synchrotron-self-Compton 
one, which indeed can account for high radiative efficiency. 

If one does not postulate the high efficiency of energy conversion
into electromagnetic radiation, then the number of
available emission mechanisms increases dramatically (e.g., Shaviv \&
Dar 1995; Panaitescu, Spada \& M\'{e}sz\'{a}ros 1999) and it becomes
almost impossible to choose only one of them on the basis of limited
observational data. At the same time, any model with low efficiency 
is problematic, as it can hardly compete with blackbody photospheric 
emission in the 10-100 keV range.

Evolution of the Lorentz factor
$\gamma_e$ (index $e$ denotes both electrons and positrons) is
governed by the acceleration and losses.
The Lorentz factor grows until the losses compensate the acceleration
term. Thus, electron acceleration is terminated at   
approximately the same Lorentz factor, $\gamma_{\rm p}$,
which is close to the value that equilibrates acceleration and
radiative losses. 

Some authors, however, prefer to consider electron injection with a
power-law spectrum, $d n_e(\gamma_e)/ dt \propto \gamma_e^{-q}$,
extending from $\gamma_{\rm min}$ to infinity (e.g., Sari et al.
1998). The spectrum is integrable ($q>2$) and hence most of
the energy is concentrated near $\gamma_{\rm min}$. In order not to
contradict our second assumption, the radiative deceleration
timescale for an electron with the Lorentz factor $\gamma_{\rm min}$
must be smaller than the GRB duration measured in the shock comoving
frame. In this case, it can be shown that the contribution of
electrons with $\gamma_e \gg \gamma_{\rm min}$ or $\gamma_e \ll
\gamma_{\rm min}$ to GRB bolometric luminosity is small. It follows
from the above consideration that the choice of electron injection
spectrum does not matter unless the question about the exact shape of
GRB spectrum is addressed; note that $\gamma_{\rm
min}$ in the approach of Sari et al. corresponds to $\gamma_{\rm
p}$ in ours and that the peak in the observed spectrum is at the 
energy determined by the electrons with $\gamma_e \sim \gamma_{\rm 
p}$.

{\it 4. The energy dissipated in the shock is efficiently 
redistributed all over the shocked gas by large-scale turbulence
and then transferred to radiating particles on a longer timescale.} 
-- We are considering a quasi-stationary problem, when the energy is 
stored in heavy particles (protons) and there is a continuous supply 
of relativistic electrons which replace those that have decelerated. 
The reason why we used a quasi-stationary formulation is that the 
bottle-neck in the energy transfer path (from protons to large-scale 
turbulence, then to electrons, then to radiation) is, most likely, at 
the first two chains, i.e., in a typical GRB, electrons are able to 
radiate at a much higher rate than they can gain energy from 
protons or large-scale turbulence.  

{\it 5. The fireball is isotropic.} -- This is a good
approximation even for a jet if it has an opening angle $\theta \gg
\Gamma^{-1}$. If the central engine generates a jet-like outflow,
then the value of $E_{\rm GRB}$ is not the actual GRB energy but
rather the energy calculated assuming isotropic emission. Therefore,
the unknown beaming factor does not affect our results.

\section{Physical parameters in GRB shocks}

For the
synchrotron-self-Compton model one may represent the luminosity of a
single highly relativistic electron 
in the following form (Ginzburg 1987, 
Rybicki \& Lightman 1979):
\begin{equation}
\label{lum-el}
{\cal L} (\gamma_e)= \frac{4}{3} \gamma_e^2 c
\left[ \sigma_T (w_{\rm m} + w_{\rm lr}) +
\int \sigma w_{\rm hr,\omega}\, d\omega \right]\, .
\end{equation}
Here $w_{\rm m}$ and $w_{\rm lr}$ are the energy densities of
magnetic field and low-energy radiation, respectively, $w_{\rm
hr,\omega}$ is the spectral energy density of high-energy radiation,
$\sigma_T$ the Thomson scattering cross-section. The boundary
between low-energy and high-energy photons is the Klein-Nishina
cut-off at $\hbar \omega \simeq m_e c^2/ \gamma_e$. Above the 
cut-off, the transport cross-section $\sigma$ is a function of 
$\gamma_e \omega$. It decreases as $(\gamma_e \omega)^{-2}$ for 
$\gamma_e \hbar \omega \gg m_e c^2$ and approaches $\sigma_T$ in the 
opposite limit.

The relative efficiency of Comptonization and synchrotron
emission is determined by the Compton $y$ parameter
(Sari, Narayan \& Piran 1996)
\begin{equation}
\label{Y}
y = \sigma_T L \int_1^{\infty} (\gamma_e^2 -1) n_e(\gamma_e)
\, d\gamma_e \simeq \gamma_{\rm p}^2 \sigma_T n_e L,
\end{equation}
where $n_e$ is the number density of electrons with the Lorentz
factors $\sim \gamma_{\rm p}$ and $L$ is the size of the emitting 
region.  For example, when electromagnetic radiation passes through a 
layer filled with relativistic electrons and the energy of incident 
photons satisfies the relation $\varepsilon \ll m_e c^2/ \gamma_e$ 
(the classical regime), the power of comptonized radiation relative 
to the power of incident radiation is given by $L_{\rm ic}/L_{\rm in} 
\equiv \tau_{\rm ic} \simeq 4y/3$. Below, we will call 
$\tau_{\rm ic}$ the optical depth for inverse Comptonization.
In general, this relation contains a numerical factor which
is different from 4/3 and depends on the geometry of a source. 

The condition $\tau_{\rm ic}=1$ (together with
$\gamma_{\rm p} < m_e c^2/ \varepsilon$) means that the source
radiates via inverse Compton scattering as much energy as via the
other emission mechanisms. If the comptonized photons may in turn
scatter off electrons in classical regime (i.e., their energy is
below the Klein-Nishina cut-off), the cascade Comptonization is
possible. For any $\tau_{\rm ic} < 1$ most of the energy will be
carried by incident radiation. Otherwise, the largest fraction of
bolometric luminosity is due to $k$ times comptonized photons, where
$k$ is the minimum number of consequent up-scatterings, which rises
an average photon energy above the Klein-Nishina cut-off.  The latter
statement is valid when the integral term in Eq. (\ref{lum-el}) is
ignored. This is possible even for $w_{\rm hr} \gg w_{\rm lr}$ thanks 
to a very rapid decrease in the transport cross-section $\sigma$. It 
should be noted, however, that this simplification may not be applied
in some cases, as discussed below.

Let us consider a relativistic shock with the Lorentz factor $\Gamma$
which radiates an energy $E_r$ during a time interval $\tau$
(in the observer frame).  The
values of $E_r$ and $\tau$ are either the total GRB energy $E_{\rm GRB}$ 
and duration  $t_{\rm
GRB}$ (in the external shock model), 
 or the energy and
duration of individual pulses (in the model of internal shocks), 
which  constitute together what we
call a GRB. As measured in the comoving frame, bolometric luminosity of
the source is $E_r/ \Gamma^2 \tau$. Given the expression for the
integral luminosity of a single particle (\ref{lum-el}), one 
easily arrives at the following relation:
\begin{equation}
\label{req2}
\frac{E_r}{\Gamma^2 \tau} = \frac{4}{3} \gamma_{\rm p}^2 \sigma_T
N_e c (w_{\rm m}+w_{\rm lr}),
\end{equation}
where $N_e$ is the total number of electrons with $\gamma_e \sim
\gamma_{\rm p}$ in the GRB shell.
The product $N_e (w_{\rm m}+w_{\rm lr})$ may be represented in the
form $n_e (W_{\rm m}+W_{\rm lr})$, where $W_{\rm m}$ and $W_{\rm lr}$
are the total energy of magnetic field and low-energy radiation in
the emitting region. Certainly, the sum $W_{\rm m}+W_{\rm lr}$ cannot
exceed a fraction of the bulk kinetic energy which is 
transferred to the particles at the shock front. According to the 
assumption {\it 2}, this fraction, $E_i$, is of the order of $E_r/ 
\Gamma$, where the factor $\Gamma$ appears because $E_i$ is measured 
in the comoving frame.

Most of the remaining factors in the right-hand side of Eq.
(\ref{req2}) can be substituted from Eq. (\ref{Y}), where
the thickness of the emitting region behind a relativistic shock
grows during the time interval $\tau$ up to $L \sim \Gamma c
\tau$ (in the comoving frame). Finally, Eq. (\ref{req2}) yields
\begin{equation}
\label{tau-cond}
\tau_{\rm ic} \ga \frac{E_r/ \Gamma}{E_i} \sim 1.
\end{equation}
Thus, there is only one possibility to satisfy all the adopted
assumptions: the observed sub-MeV emission from GRBs should be well
above the Klein-Nishina cut-off for electrons in the emitting region
in order not to generate too much ultra-hard radiation via the inverse
Compton process.

To specify this requirement quantitatively, it is
necessary to know the low-energy portion of the GRB spectrum. If it is a
power-law with the index $\alpha$ ($\omega I_{\omega} \propto
\omega^\alpha$), then according to assumption {\it 1} we have
\begin{equation}
\label{gamma-lim}
\left( \frac{\gamma_{\rm p} \varepsilon_p}
{\Gamma m_e c^2} \right)^{\alpha^{\prime}} \ga \tau_{\rm ic}
\qquad \Rightarrow \qquad \gamma_{\rm p} \ga 2.5\,
\Gamma \tau_{\rm ic}^{1/ \alpha^{\prime}}\, ,
\end{equation}
where we use the fact that the observed spectrum of an average GRB
has a maximum at $\varepsilon_p \sim 200$ keV. In Eq.
(\ref{gamma-lim}) $\alpha^{\prime} = \mbox{min}\, \{\alpha,2\}$: for
spectra rising steeper than $\omega^2$ the contribution of photons
near $\varepsilon_p$ to the inverse Compton cooling rate dominates,
despite a much smaller transport cross-section $\sigma$.

When synchrotron-self-Compton mechanism is the main one operating in
the source, the maximum of the observed GRB spectrum is at
\begin{equation}
\label{spec-max}
\varepsilon_p \sim 0.4\, \Gamma \left( \frac{4}{3} \gamma_{\rm p}^2
\right)^{k+1} \frac{\hbar e B}{m_e c}\, ,
\end{equation}
where $k$ is the number of Comptonization cascades; $k=0$ corresponds
to the synchrotron radiation. The magnetic field strength may be
estimated from Eq. (\ref{req2}), where one has to establish the
relation between $w_{\rm m}$ and $w_{\rm lr}$. Given the expression
(\ref{lum-el}) for the
synchrotron luminosity ($w_{\rm lr}=w_{\rm hr}=0$) of a single
particle, the energy density of synchrotron radiation is
\begin{equation}
w_{\rm sy} \simeq \frac{4}{3} \gamma_{\rm p}^2 \sigma_T c
\frac{B^2}{8\pi} n_e \frac{L}{c} = \tau_{\rm ic} w_{\rm m}.
\end{equation}
This relation follows directly from Eq. (\ref{lum-el}). 
It is well-known that each subsequent cascade of Comptonization also 
leads to a $\tau_{\rm ic}$ times increase in the energy density of 
comptonized photons. Finally, one obtains 
\begin{equation} 
\label{B-w} 
w_{\rm lr} + w_{\rm m} = \sum_{j=0}^k \tau_{\rm ic}^j \frac{B^2}{8\pi} 
\simeq \tau_{\rm ic}^k \frac{B^2}{8\pi}\, , 
\end{equation} 
where the approximate equality is for $\tau_{\rm ic} \gg 1$.

After some algebra, i.e., deriving the magnetic field strength from
Eq. (\ref{B-w}) with the help of the relation (\ref{req2}) and
substituting $B$ in Eq. (\ref{spec-max}), one has the following
result
\begin{equation}
\label{maxGamma-e}
\gamma_{\rm p} \simeq \frac{\sqrt{3}}{2}
\tau_{\rm ic}^{\frac{1}{4}} \Gamma^{\frac{1}{k+1}}
\left( \frac{2.5 m_e c \varepsilon_p}{e \hbar} \right)^
{\frac{1}{2(k+1)}}
\left( \frac{c^3 \tau^3}{2\, E_r} \right)^{\frac{1}{4(k+1)}},
\end{equation}
which should be compared with the requirement (\ref{gamma-lim}).
Thus,
\begin{equation}
\label{tau-win}
\tau_{\rm ic}^{\frac{1}{\alpha^{\prime}}-\frac{1}{4}} 
\la \frac{\sqrt{3} \varepsilon_p}{2 m_e c^2}
\Gamma^{-\frac{k}{k+1}}
\left( \frac{2.5 m_e c \varepsilon_p}{e \hbar} \right)^
{\frac{1}{2(k+1)}}
\left( \frac{c^3 \tau^3}{2\, E_r} \right)^{\frac{1}{4(k+1)}}.
\end{equation}
Because we know that $\alpha^{\prime} \leq 2$ and $\tau_{\rm ic} \ga
1$ (see Eq. (\ref{tau-cond})), the conclusion is straightforward:
$k=0$, except the extreme case
\begin{equation}
\begin{array}{l}
\displaystyle
\Gamma \la \frac{3 \varepsilon_p^2}{4 m_e^2 c^4}
\left( \frac{2.5 m_e c \varepsilon_p}{e \hbar} \right)^{1/2}
\left( \frac{c^3 \tau^3}{2\, E_r} \right)^{1/4} \\
\displaystyle
\phantom{\Gamma }
\sim 25 \frac{t_1^{3/4}}{E_{52}^{1/4}}
\left( \frac{\tau}{t_{\rm GRB}} \right)^{1/2}.
\end{array}
\end{equation}
That is, the radiation which is
observed from GRBs in the sub-MeV spectral range is generated by
a synchrotron mechanism.

Substituting $k=0$ into Eq. (\ref{tau-win}) we find that the value of
$\tau_{\rm ic}$ must fit in the following window
\begin{equation}
\label{tau-lim}
1 \la
\tau_{\rm ic}^{\frac{1}{\alpha^{\prime}}-\frac{1}{4}} \la
300 \left( \frac{\varepsilon_p}{m_e c^2} \right)^{\frac{3}{2}}
\left( \frac{ t_1^3}{E_{52}} \right)^{\frac{1}{4}}
\left( \frac{\tau}{t_{\rm GRB}} \right)^{\frac{1}{2}},
\end{equation}
where $t_1$ is the duration of GRB in units of 10~s, $E_{52}$ the burst
energy in units of $10^{52}$~erg, and for simplicity the
luminosity in peaks, $E_r/\tau$, is assumed to be of the order of
the average luminosity, $E_{\rm GRB}/t_{\rm GRB}$.
Equation (\ref{tau-lim}) gives a lot of freedom for $\alpha^{\prime}
\simeq 2$.  However, we suppose that the spectrum is much flatter,
with $\alpha$ not larger than unity, which is true for the majority of
GRBs (Tavani 1996). In this case, the result of our analysis
becomes more definite: the optical depth for inverse Comptonization
is between 1 and 30, approaching 100 in extreme cases. 
Consequently, the magnetic
field strength $B$ is rather close to its equipartition value
(here we mean the equipartition with the energy stored in protons),
\begin{equation}
\label{Bequ}
B \simeq \left[ \frac{2\, E_r}
{(\Gamma^2 c \tau)^3 \tau_{\rm ic}} \right]^{1/2} \sim
\frac{E_{52}^{1/2}}{t_1^{3/2} \tau_{\rm ic}^{1/2}} 
\frac{t_{\rm GRB}}{\tau} \frac{10^9}{\Gamma^3}\, \mbox{G}.
\end{equation}
An estimate of $\gamma_{\rm p}$ may be obtained from Eq.
(\ref{maxGamma-e}):
\begin{equation}
\gamma_{\rm p} \simeq 200\, \tau_{\rm ic}^{1/4} \Gamma
\left( \frac{t_1^3}{E_{52}} \right)^{\frac{1}{4}}
\left( \frac{\tau}{t_{\rm GRB}} \right)^{\frac{1}{2}}.
\end{equation}

A significant fraction of the total GRB energy is converted via
inverse Compton scattering into ultra-hard emission.
The typical energy of comptonized photons corresponds to the
Klein-Nishina cut-off at
\begin{equation}
\label{ic-max}
\varepsilon_{\rm ic} \sim \Gamma \gamma_{\rm p} m_e c^2 \sim
10^{-4}\, \Gamma^2
\left( \frac{t_1^3}{E_{52}} \right)^{\frac{1}{4}}
\left( \frac{\tau}{t_{\rm GRB}} \right)^{\frac{1}{2}} \mbox{TeV} 
\end{equation}
in the observer frame.
The ratio of energy carried by the ultra-hard radiation to the
energy carried by the sub-MeV radiation is
\begin{equation}
\begin{array}{l}
\displaystyle 
\delta E_{\rm ic} \simeq
\left( \frac{\Gamma m_e c^2}{\gamma_{\rm p} \varepsilon_p}
\right)^{\alpha} \tau_{\rm ic} \\
\displaystyle 
\phantom{\delta E_{\rm ic} }
\ga \left[ 1.3 \cdot 10^{-2}
\left( \frac{E_{52}}{t_1^3} \right)^{1/4}
\left( \frac{t_{\rm GRB}}{\tau} \right)^{1/2}
\right]^{\alpha}.
\end{array}
\end{equation}
For sufficiently flat spectra ($\alpha < 1$), which are considered
here, the relative energy content of ultra-hard radiation is always
non-negligible. In the external shock model one has $\delta E_{\rm
ic} \ga 1\%$. Moreover, in the model of internal shocks, in which
$\tau \sim 1-10$~ms (presumably), the value of $\delta E_{\rm ic}$
should be close to its limiting value of unity set by the assumption {\it
1}, independently on the spectral index $\alpha$.

\section{Implications of the analysis}

The principle outcome of the above analysis is that the observed
sub-MeV emission in GRBs is generated by the synchrotron
mechanism. However, the limits given in Eq. (\ref{tau-lim}) become
mutually exclusive if
\begin{equation}
\label{death}
\left( \frac{\varepsilon_p}{m_e c^2} \right)^{3}
\left( \frac{ t_1^3}{E_{52}} \right)^{\frac{1}{2}}
\frac{\tau}{t_{\rm GRB}} < 10^{-5}\, .
\end{equation}
The bursts having parameters in
this domain can radiate only a minor part of the available energy in
the sub-MeV spectral range. Therefore, Eq. (\ref{death}) defines the
 synchrotron-self-Compton (SSC) constraint for Gamma-Ray Bursts.

There are two implications of the SSC constraint. It may
be treated as a lower limit on GRB duration (the external shock
model, $\tau = t_{\rm GRB}$):
\begin{equation}
t_{\rm GRB} > 5\cdot 10^{-3}\, E_{52}^{1/3}
\left( \frac{m_e c^2}{\varepsilon_p} \right)^{2} \mbox{s}
\sim 0.03\, \mbox{s}.
\end{equation}
On the other hand, Eq. (\ref{death}) determines the shortest possible
variability timescale $\tau$ in the model of internal shocks:
\begin{equation}
\tau > 10^{-5}
\left( \frac{m_e c^2}{\varepsilon_p} \right)^{3}
\left( \frac{E_{52}}{ t_1^3} \right)^{\frac{1}{2}}
t_{\rm GRB} \sim 10^{-3}\, \mbox{s}.
\end{equation}
A variability significantly shorter than the limiting duration 1 ms
-- if observed in GRB lightcurves -- favours propagation effects or
coherent emission mechanisms as a primary explanation for short-time
pulsations in GRBs.

In order not to contradict assumption {\it 2}, the deceleration
timescale for an electron with the Lorentz factor $\gamma_{\rm p}$
must be smaller than the GRB duration measured in the shock comoving
frame. Since we know the relations between $B$, $\gamma_{\rm p}$
and the parameters characterizing a GRB ($E_{\rm GRB}$, $t_{\rm
GRB}$, $\tau$ and $\Gamma$), it is possible to use this constraint
to set limit on $\Gamma$:
\begin{equation}
\begin{array}{l}
\displaystyle \Gamma \la
\left( \frac{\sigma_T}{4\pi \sqrt{3}} \right)^{\frac{1}{4}}
\left( \frac{2.5 \varepsilon_p}{e\hbar m_e c} \right)^{\frac{1}{8}}
\left( \frac{2 E_{\rm GRB}}{\tau_{\rm ic} c^3 t_{\rm GRB}^3}
\right)^{\frac{3}{16}}
\left( \frac{t_{\rm GRB}}{\tau} \right)^{\frac{1}{8}}
t_{\rm GRB}^{\frac{1}{4}}\\
\displaystyle \phantom{\Gamma}
\simeq 1.2 \cdot 10^3\, \frac{E_{52}^{3/16}}{\tau_{\rm ic}^{3/16}
t_1^{5/16}} \left( \frac{t_{\rm GRB}}{\tau} \right)^{\frac{1}{8}}.
\end{array}
\end{equation}
Exactly the same limit may be derived from the requirement that the
total energy of the accelerated electrons does not exceed the GRB 
energy.  It follows from the above analysis that the accelerated 
electrons, in the general case, are not in equipartition with 
the magnetic field.

An important question to be addressed in the theory of synchrotron
emission is why the value of $\varepsilon_p$ is nearly invariant.
This problem may be reformulated as a question regarding the electron
acceleration mechanism. The simplest reasonable suggestion is that
$\gamma_{\rm p}$ is a function of magnetic field strength $B$,
which is given by Eq. (\ref{Bequ}). To eliminate 
$\Gamma$-dependence (the strongest one in Eq. (\ref{spec-max})),
the acceleration mechanism should
provide $\gamma_{\rm p} \propto B^{-1/3}$. In this case,
$\varepsilon_p$ preserves a weak dependence on $E_r$ and $\tau$,
$\varepsilon_p \propto E_r^{1/6} \tau^{-1/2}$, where the dependence
on $\tau$ is likely to be hidden by a stronger effect of
the cosmological time dilation.

Several per cent (a few tens per cent is a more likely
figure) of the total GRB energy must be converted in the ultra-hard
spectral domain via the inverse Compton scattering. The location of 
the maximum in the spectrum of comptonized radiation is 
model-dependent.  In the model of internal shocks, the maximum (see 
Eq. (\ref{ic-max})) may be at energies as low as $10^{-6}\, 
\Gamma^2$~TeV$\sim 10-100$~GeV, so that the number fluence of 
ultra-hard photons should be of the order of $10^3$~km$^{-2}$ even 
for the weakest bursts. However, the uncertainty in theoretical 
predictions is large: the maximum in the spectrum of inverse Compton 
radiation may be located at $\sim 10^3$ TeV. On the other hand, in 
the external shock model, the location of the maximum is more 
definite, namely, $\sim 10^{-4}\, \Gamma^2$~TeV$\sim 1-100$~TeV. In 
this case the ultra-hard emission is inaccessible for direct 
observation because photons above several hundred GeV are strongly 
absorbed by the infrared background radiation (Primack et al. 1999), 
and the observed fluence below 1~TeV is determined by the spectral 
slope of comptonized radiation in TeV energy range. To date, there is 
only one indication of sub-TeV emission accompanying GRBs (Atkins et 
al.  2000).

According to our estimates, in both models, the GRB source may be
optically thick for the two-photon absorption of the highest energy
quanta.  Qualitative results of the two-photon absorption are the
following (details will be discussed elsewhere).  Each of the
absorbed photons produces an electron-positron pair, in which the
daughter particles have roughly equal energies.  Electrons and
positrons have a larger interaction cross-section than photons have. 
Therefore, 
 if the optical depth for interaction of initial ultra-hard
quanta is larger than unity, the same is true for daughter electrons
(positrons). The first photon scattered off by one of these energetic
particles takes away about a half of its energy, and in turn may be
absorbed. Step by step, the radiation becomes softer until the
absorption threshold is reached.  It is this spectral domain where
the bulk of energy initially contained in the ultra-hard radiation is
concentrated. If the absorption threshold is below 1 TeV, then the
reprocessed radiation may reach the Earth. A similar picture arises
if one considers the reprocessing of the ultra-hard photons via
interaction with soft GRB radiation scattered in interstellar medium
(Derishev, Kocharovsky \& Kocharovsky 2000).

The result $\tau_{\rm ic} \sim 1 - 30$ seems
surprising since $\tau_{\rm ic}$ is an integral characteristic of the
emitting region, while it depends on $\gamma_{\rm p}$, which is
determined by the balance between acceleration and losses and,
therefore, is defined by local conditions. Such a fine
tuning in the external shock model may result from the following.
We have already noted that the statement $\tau_{\rm ic}
= 1$ means that the energy densities of the magnetic field and
synchrotron radiation are equal in the emitting region. Since this
region expands with a relativistic velocity, a large fraction of
newly born synchrotron photons cannot escape it until a GRB shell as
a whole is decelerated to a Lorentz factor much smaller than the
initial one.  During the initial deceleration stage the synchrotron
radiation is accumulated to the point at which its energy density 
becomes nearly equal to the thermal energy density of plasma. If 
strong turbulence in a GRB shell rapidly builds up a magnetic field 
of a strength close to the equipartition value, then the approximate 
correspondence between the energy density of magnetic field and that
of synchrotron radiation is established automatically, thus explaining
why $\tau_{\rm ic} \sim 1$. In the model of internal shocks, it is
usually assumed that the Lorentz factor of successive shells differs
by a factor of few. Nearly all generated radiation remains within
the emitting region and the condition $\tau_{\rm ic} \sim 1$ is
satisfied automatically.

The situation when $\tau_{\rm ic} >1$ and, at the same time,
the synchrotron emission is the most efficient cooling mechanism in
the portion of electron distribution with the highest luminosity
yields the largest flexibility in spectral shapes of
synchrotron radiation. So far, we considered the
low-energy spectral slope as an independent parameter, assuming only
that $\alpha \leq 2$. The complete theory of synchrotron-self-Compton
emission, however, should give a definite prediction of $\alpha$.

In the simplest approximation only the synchrotron losses are taken
into account, so that partially decelerated electrons form the
following distribution below $\gamma_{\rm p}$:
\begin{equation}
\label{dist1}
n_e(\gamma_e) = \frac{m_e c^2}{{\cal L} (\gamma_e)}
\frac{dn_e}{dt} = n_e \frac{\gamma_{\rm p}}{\gamma_e^2}.
\end{equation}
For a power-law electron distribution, $n_e(\gamma_e) \propto
\gamma_e^q$, the synchrotron spectrum is also a power-law, $\omega
I_\omega \propto \omega^{\frac{q+3}{2}}$. Distribution (\ref{dist1})
gives the well-known result, $\omega I_\omega \propto \omega^{1/2}$.

A different approximation works well when the integral term in Eq.
(\ref{lum-el}) dominates. In this case, the main contribution to the
integral is from photons near the Klein-Nishina cut-off whose
frequency scales as $\gamma_e^{-1}$. Looking for a power-law
solution, we find that ${\cal L} \propto \gamma_e^{2-\alpha}$ if
$\omega I_\omega \propto \omega^{\alpha}$, and that $n_e(\gamma_e)
\propto \gamma_e^{\alpha-2}$ consequently. Thus,
$\alpha=(\alpha+1)/2$ and the final result is
\begin{equation}
\label{dist2}
n_e(\gamma_e) = \frac{n_e}{\gamma_e}
\qquad \mbox{and} \qquad I_\omega = const.
\end{equation}

As follows from the above analysis, the GRB sources are at the
border between these two limiting cases. Namely, the
synchrotron losses dominate for electrons with $\gamma_e \sim
\gamma_{\rm p}$, but the inverse Compton losses become dominant
for less energetic electrons
because the Klein-Nishina cut-off is shifted to higher photon
energies, giving greater $w_{\rm lr}$. The fact that all
GRBs belong to the transition zone between two limiting cases makes
theoretical analysis more complicated, but also allows for a wider
range of solutions.

Let us consider one example to illustrate this.
Suppose that electrons with the Lorentz factors
between $\gamma_{\rm ic}$ and $\gamma_{\rm p}$ lose their energy
mainly due to synchrotron radiation. The electrons in this range
form the distribution (\ref{dist1}) and produce the corresponding
spectrum $\omega I_\omega \propto \omega^{1/2}$, which extends from
$\varepsilon_p$ down to $(\gamma_{\rm ic}/\gamma_{\rm p})^2
\varepsilon_p$. If the Klein-Nishina cut-off, $\varepsilon_{\rm
k-n}$, lies somewhere between $(\gamma_{\rm ic}/\gamma_{\rm p})^2
\varepsilon_p$ and $(\gamma_{\rm ic}/\gamma_{\rm p}) \varepsilon_p$
for $\gamma_e \sim \gamma_{\rm p}$, then the relative importance of
the integral term in Eq. (\ref{lum-el}) grows until $\gamma_e <
\gamma_{\rm k-n} = \varepsilon_{\rm k-n}/\varepsilon_p$. Electron
distribution between $\gamma_{\rm k-n}$ and $\gamma_{\rm ic}$ is
determined by the spectral slope in the region just below
$\varepsilon_p$, where, in this case, the spectral index is 
$\alpha=1/2$. Hence, electron distribution is given by
$n_e(\gamma_e) \propto \gamma_e^{-3/2}$ (see the discussion of Eq.
(\ref{dist2})) and the synchrotron spectrum generated by
these electrons has an index $\alpha=3/4$ between energies
$(\gamma_{\rm ic}/\gamma_{\rm p})^2 \varepsilon_p$ and
$(\gamma_{\rm k-n}/\gamma_{\rm p})^2 \varepsilon_p$. At the lowest
energies, the spectral index is again $\alpha=1/2$, since the inverse
Compton losses saturate when the Klein-Nishina cut-off is shifted
above $\varepsilon_p$. 

The GRB spectrum below $\varepsilon_p$ is a broken power-law 
consisting of two portions with the spectral index $1/2$ separated by 
a steeper part with $\alpha=3/4$. Therefore, in addition to the
well-known cooling break in the synchrotron spectrum there are two 
others -- the lower-energy Klein-Nishina break, below which the ratio 
of synchrotron to inverse Compton losses levels off, and the 
higher-energy Compton break, above which synchrotron losses overcome 
inverse Compton ones. Locations of all spectral breaks are 
time-dependent both during the main pulse and at the afterglow stage, 
and if one is limited to a narrow-band observations, the transition 
from one spectral slope to another could be interpreted as a change 
in the decline rate of a lightcurve. A similar effect may appear for 
a bolometric lightcurve when the balance between synchrotron 
and inverse Compton emissivities is shifted in favor of the latter.

One more effect may alter the appearance of the low-energy part of the
GRB spectrum -- the synchrotron self-absorption. With the help of
expressions for $\gamma_{\rm p}$ and $\tau_{\rm ic}$, it is not
difficult to estimate the self-absorption threshold energy
(Zheleznyakov 1996). It takes the highest value in the model of
internal shocks, but even in this model the threshold is hardly above
500 eV in the observer frame.

\section{Conclusions}

The analysis in this paper relies to a large extent upon several
assumptions which have not received yet direct observational
confirmation. However, we put forward a number of reasons why these
assumptions are consistent with the known properties
of GRBs. Our investigation reveals that, in the range of
parameters typical for the detected bursts, there is no other
possibility that accounts for the sub-MeV radiation 
than synchrotron emission of electrons accelerated in relativistic 
shocks. On this basis, we make several definite predictions of
physical parameters in relativistic shocks of GRB origin.

One of the most important conclusions is that physical conditions in
GRB emitting regions are qualitatively similar. Namely, there is not 
so much freedom in the magnetic field strength, which is close to its 
equipartition value, and in the Lorentz factor of accelerated 
electrons (Eq. \ref{maxGamma-e}). The main uncertainty in the energy 
of comptonized photons arises from differences between external and 
internal shock models and from poor knowledge of the bulk Lorentz 
factor. The inverse Compton losses always play a significant 
role, so that the ultra-hard emission above 100 GeV contains from 
several to tens per cent of the total GRB energy. 

We demonstrate that a steady-state electron distribution 
consistent with the Compton losses may produce different spectral 
indices, ranging from the widely accepted 1/2 to 1. In many GRBs, the 
competition between inverse Compton and synchrotron losses should 
produce rather complex (power-laws with several breaks) spectra in 
the X-ray and soft $\gamma$-ray ranges. The time evolution of 
spectral breaks, as well as transition from predominantly synchrotron 
losses to predominantly inverse Compton losses, may account for the 
changes in the decline rate observed in the lightcurves of several 
GRB afterglows. 

The prevalence of a synchrotron emission mechanism in GRBs imposes 
the strict limitation on the burst parameters, which may be called 
a synchrotron-self-Compton constraint. The physical meaning of 
this constraint is that there cannot be radiatively efficient 
bursts with arbitrary small duration or arbitrary fast variability. 
The requirement of high radiative efficiency also limits the Lorentz 
factor of GRB fireballs to a value not larger than $\sim 10^{3} - 
10^{4}$, depending on the model (external or internal shocks, 
respectively). So, the range of admissible Lorentz factors in the 
GRB case is limited from both sides. 

\section*{Acknowledgements}

This work has been supported by Russian Foundation for Basic Research
(the project 99-02-18244) and by the Russian Academy of Science
through a grant for young scientists.


\begin{thebibliography}{}

\bibitem{rotse} Akerlof, C. {\it et al.}, 1999,
Nature 398, 400.

\bibitem{milagro} Atkins, R., {\it et al.}, 2000, ApJ 533, L119.

\bibitem{gamma} Baring, M.G., Harding, A.K., 1995, Adv. Space Res.
V.15, N.5, 153.

\bibitem{polar1} Covino, S. {\it et al.}, 1999, A\&A 348, L1.

\bibitem{apj} Derishev, E.V., Kocharovsky, V.V., Kocharovsky, Vl.V.,
1999, ApJ 521, 640.

\bibitem{hgrb} Derishev, E.V., Kocharovsky, V.V., Kocharovsky, Vl.V.,
2000, Proceedings of the 5th Huntsville Gamma-Ray Burst
Symposium, 18-22 October 1999, Huntsville, USA, p. 460.

\bibitem{4} Frail, D.A. {\it et al.}, 1997, Nature 389, 261.

\bibitem{decay1} Galama, T.J. {\it et al.},
1998, ApJ 497, L13.

\bibitem{Ginz} Ginzburg, V.L., {\it Theoretical physics and
astrophysics}, Nauka, Moscow, 1987.

\bibitem{sy-ic} M\'{e}sz\'{a}ros, P., Laguna, P., Rees, M.J., 1993, 
ApJ 415, 181.

\bibitem{jet} M\'{e}sz\'{a}ros, P., Rees, M.J., Wijers, R.A.M.J.,
1999, New Astr. 4, 303.

\bibitem{3} Metzger, M.R. {\it et al.}, 1997, Nature 387, 878.

\bibitem{r-aft} Paczy\'{n}ski, B., Rhoads J.E., 1993, ApJ
418, L5.

\bibitem{mech2} Panaitescu, A., Spada, M., M\'{e}sz\'{a}ros, P., 1999,
ApJ 522, L105.

\bibitem{rev} Piran, T., 1999, Phys. Rep. 314, 575.

\bibitem{TevAbs} Primack, J.R., Bullock, J.S., Somerville, R.S., 
MacMinn, D., 1999, Astropart. Phys. 11, 93.

\bibitem{IntSh} Rees, M.J., M\'{e}sz\'{a}ros, P., 1992, MNRAS 258,
P41.

\bibitem{ExtSh} Rees, M.J., M\'{e}sz\'{a}ros, P., 1994, ApJ 430, L93.

\bibitem{2} Reichart, D.E., 1997, ApJ 485, L57.

\bibitem{GRBen} Reichart, D.E., M\'{e}sz\'{a}ros, P., 1997, ApJ 
483, 597.

\bibitem{rbl} Rybicki, G.B., Lightman, A.P., {\it Radiative Processes 
in Astrophysics}, 1979.

\bibitem{sari2} Sari, R., Narayan, R., Piran, T., 1996, ApJ 473, 204.

\bibitem{sari} Sari, R., Piran, T., Narayan, R., 1998, ApJ 497, L17.

\bibitem{mech1} Shaviv, N., Dar, A., 1995, MNRAS 277, 287.

\bibitem{syncS} Tavani, M., 1996, ApJ 466, 768.

\bibitem{23} Waxman, E., 1997, ApJ 485, L5.

\bibitem{polar2} Wijers, R.A.M.J. {\it et al.}, 1999, ApJ 523, L33.

\bibitem{Zh} Zheleznyakov, V.V., {\it Radiation in
astrophysical plasmas}, Kluwer, Dordrecht/Boston/London, 1996.


\end{thebibliography}
\end{document}